\documentclass[aps,twocolumn,showpacs,amsmath,amssymb,pra,superscriptaddress,floatfix,longbibliography]{revtex4-1}

\usepackage{times}
\usepackage{t1enc}
\usepackage{subfigure}
\usepackage{graphicx}
\usepackage{epstopdf}
\usepackage{amsmath}
\usepackage{dcolumn}
\usepackage{xcolor}
\usepackage{soul}
\usepackage{hyperref}

\hypersetup{
    colorlinks=true,
    linkcolor=blue,
    filecolor=magenta,      
    urlcolor=blue,
    citecolor = blue,
    }

\begin{document}


\title{Emission photon statistics in collectively interacting dipole atom arrays in the low-intensity limit}

\author{Deepak A.~Suresh}
\affiliation{Department of Physics and Astronomy, Purdue University, West Lafayette,
Indiana 47907, USA}

\author{F.~Robicheaux}
\email{robichf@purdue.edu}
\affiliation{Department of Physics and Astronomy, Purdue University, West Lafayette,
Indiana 47907, USA}
\affiliation{Purdue Quantum Science and Engineering Institute, Purdue
University, West Lafayette, Indiana 47907, USA}




\date{\today}

\begin{abstract}

We investigate the photon statistics of light emitted from a system of collectively interacting dipoles in the low-intensity regime, incorporating double-excitation states to capture beyond-single-excitation effects. By analyzing the eigenstates of the double-excitation manifold, we establish their connection to the accessible single-excitation eigenmodes and investigate the role of decay rates in shaping the zero time delay photon correlation function $g^{(2)}(\tau = 0)$ under different detection schemes. 
The photon emission statistics can be arbitrarily controlled by interfering two beams of light that selectively address orthogonal eigenmodes. This can act as a tunable nonlinearity that enables both enhancement or suppression of two-photon emission.

\end{abstract}


\maketitle

\section{Introduction\label{sec:intro}}

Making photons interact has been a long-standing goal in optics. 
The regime of non-linear effects at low intensities, where individual photons interact strongly with one another, can be called 'Quantum nonlinear optics' \cite{cv2014}.
It holds great potential for applications in optical transistors, non-linear optical switches \cite{m2010}, quantum information and communication \cite{k2008}, and metrology using non-classical fields \cite{ms2004,gl2011}.
The scattering of two photons with the help of interaction with other systems has been explored in Refs. \cite{rw2017,lc2012,yr2008,poddubny1_2019,sf2007}

At the single-photon level, collective interactions in free space have had many theoretical proposals and experimental implementations of applications in coherent control \cite{dicke,bettles,re1971,fhm1973,gfp1976,s2009,mss2014,pbj2014,bbl2016,bzb2016,psr2017,jbs2018,zoller,chs2003,ymg2013,mmm2018,jrp2017,bpp2020,ggv2018,wzc2015}. The photon statistics associated with these collective effects, especially in systems with many emitters, have also been investigated \cite{wrv2020,agp2011,tw2009,mh2010,gsd2005}.
This interaction with emitters has been harnessed to make individual photons interact with each other, especially in the more controllable case of emitters coupled to waveguides \cite{ph2020,sp2023,zb2013,ah2019,zm2019,zo2020,zy2020,p2020,ms2021,pp2023,ff2023,je2024,cs2023}.

In the case of atom-light interactions with high intensities, many of the atoms are simultaneously excited, sometimes even reaching full inversion. Superradiance and subradiance can be seen in the emission of photons. Ref. \cite{ma2022,FR2021} have studied the correlations in the photons emitted in such systems. They showed that bunching is a characteristic of the emission in superradiant systems in this regime of near-total inversion. Alternatively, will these associations be valid in the opposite regime, where the intensity is low enough that there are only one or two excitations on average? The process of building coherences for the enhanced/suppressed emission is quite different in the two cases. 

When atoms interact with the incident light, there are two types of contributions that give rise to photon correlations. 
The first is due to the interference of the incoming driving light and the emitted light, which is analogous to the $g^{(2)}$ in the forward direction in systems such as atom arrays. In the low-intensity limit, this effect primarily comes from the single-excitation states of the ensemble, which has a population proportional to the intensity ($\propto \Omega^2$). The terms from the double-excitation states will be proportional to the square of intensity ($\propto \Omega^4$) and will not have a significant contribution.

Alternatively, in the backward direction, the correlations arise only because of the light emitted by the ensemble.
These correlations arise purely because of the interactions between the atoms in the ensemble. The second-order correlation between two emitted photons no longer depends only on the single-excitation states, but also on the doubly-excited states. Hence, we will include and focus on how the doubly-excited states evolve and contribute to the dynamics of the second-order correlation of the photon emission. We study the correlations only in the emitted light, without including the incoming laser field, which is analogous to studying the backward scattering of light. 

In this paper, we explore how the correlations in the emitted light are affected by the decay rate of the system. 
The study focuses on a system consisting of collectively interacting dipole atoms arranged in sub-wavelength arrays. The atoms are excited by constant incident light with weak intensity, and the system is driven to steady state. 
We particularly emphasize investigating the zero-time-delay correlation function $g^{(2)}(\tau = 0)$ to determine the emission statistics. It describes the probability of two photons being emitted together, normalized by their independent emission. When the photons emitted hold no correlations, it will have $g^{(2)}(0) = 1$. If the photons are more prone to being emitted together, it is called bunched emission and will have a $g^{(2)}(0) > 1$. The opposite is called anti-bunching and will have a $g^{(2)}(0) < 1$.

We study these dynamics using a density matrix formalism using both the single and double-excitation states of the ensemble. 
We study how the zero-time-delay $g^{(2)}(0)$, which pertains to the emission of two photons together, depends quantitatively on the decay rates of both the single and double-excitation eigenmodes. Although double-excitation states have been studied in other contexts \cite{poddubny1_2019,uk2024}, our study will focus on the double-excitation states as a consequence of working beyond the edge of the low-intensity/single-excitation regime.

We also observed that interfering two coherent beams provides a mechanism for controlling the photon statistics of the emitted light. By selectively detecting light from a single eigenmode, we can tune the $g^{(2)}(0)$ between antibunching and bunching by adjusting the relative phase of the beams. A source capable of real-time modulation of its correlation properties without altering intensity can be used as an interesting building block for novel quantum light sources. This can also be utilized to suppress two-photon emission into a desired mode. References \cite{cs2023,je2024,sc2024} have also studied different methods of tailoring the correlations of light using interferences from multiple emitters.


The formalism used for calculations is detailed in Sec. \ref{sec:methods}. Section \ref{sec:DoubleEx} investigates the properties of the double-excitation states and how they connect to the single-excitation manifold. 
Section \ref{sec:SingleMode} investigates the initial time correlation, $g^{(2)}(0)$, of the emitted light when the atoms are excited by a single eigenmode.
The effects of interference of two different modes is explored in Sec. \ref{sec:DoubleMode}

\section{Methods\label{sec:methods}}
We consider N atoms and assume each to be a two-level system. The atoms are arranged in 1D or 2D arrays with separation $(d)$ less than the wavelength of the resonant light $(\lambda)$. The dynamics of the system are calculated using the density matrix formalism.

The low-intensity limit is often assumed when performing calculations for collective interaction dynamics. There can be only a single excitation in the system at any time, which means that only the single-excitation states have to be considered in the density matrix. The $g^{(2)}$ correlations, which describe the correlation between the emission of two photons, also require the doubly-excited states to be considered. This increases the total number of states (of the order $N^2$), which considerably reduces the total number of atoms that can be accurately simulated.


\begin{figure}
    \centering
    \includegraphics[width=0.99\linewidth]{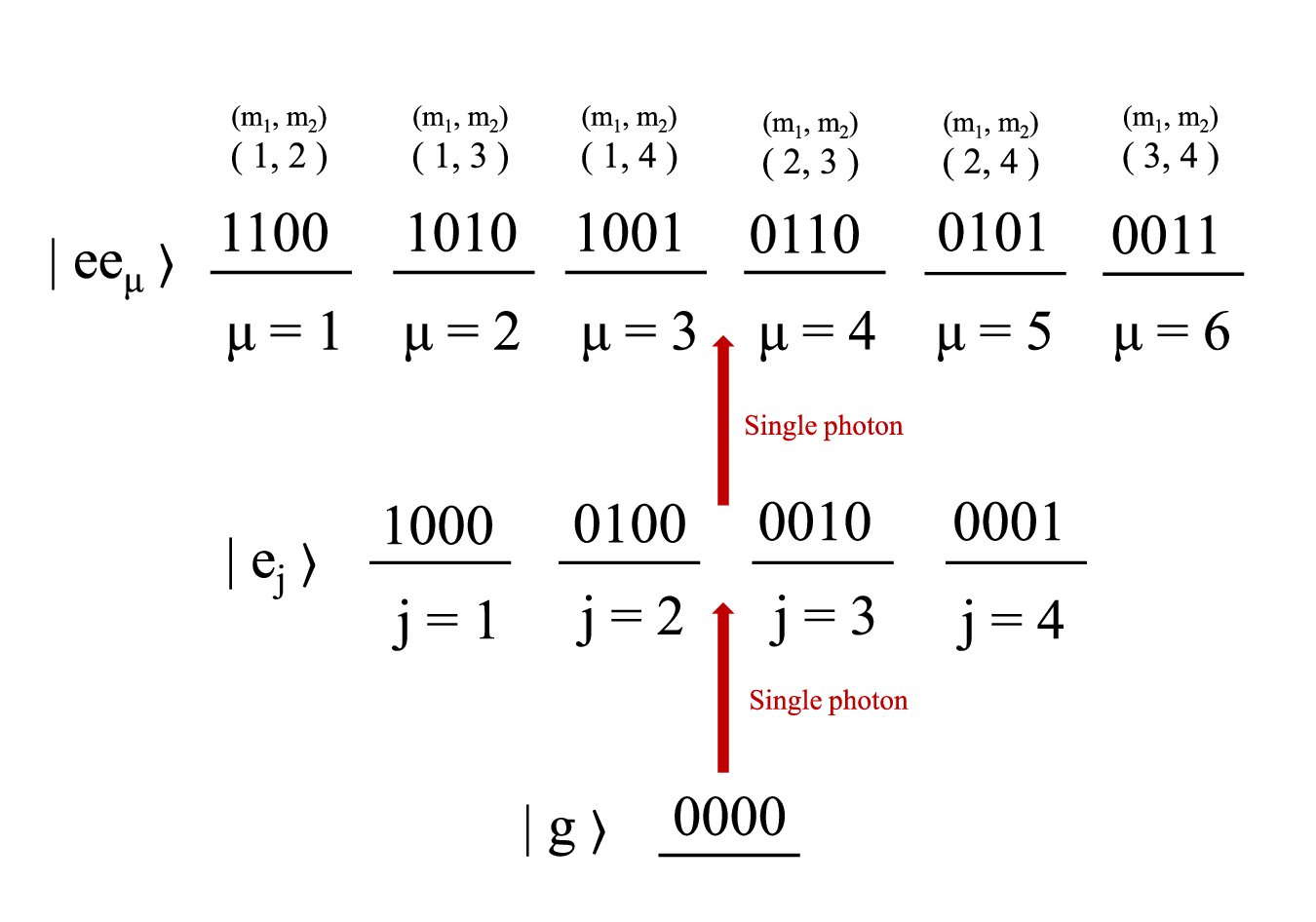}
    \caption{An example schematic of the atomic energy levels for $N = 4$ atoms. 0 and 1 represent the atoms being in the ground and excited state, respectively. $|g\rangle$ represents the collective ground state. $|e_j\rangle$ represents the single-excitation states and $|ee_{\mu}\rangle$ represents the double-excitation states.}
    \label{fig:schematic}
\end{figure}

The raising and lowering operators of the $j^{th}$ atom are represented by $\hat{\sigma}^+_j$ and $\hat{\sigma}^-_j$ respectively.
The state in which all atoms are in the ground state is represented by $|g\rangle$ and the states in which only the atom $'j'$ is excited will be represented by $|e_j\rangle = \hat{\sigma}^+_j|g\rangle$. The doubly-excited state will be represented by $|ee_{\mu}\rangle = \hat{\sigma}^+_{m_1}\hat{\sigma}^+_{m_2}|g\rangle$ and the index $\mu = (m_1,m_2)$ represents the atoms $m_1$ and $m_2$ being excited. Since $m_1 < m_2$, the index $\mu$ goes from $1$ to $N(N-1)/2$. An example schematic of the energy level structure for $N = 4$ atoms has been depicted in Fig. \ref{fig:schematic}. Hence the density matrix will be represented by 
\begin{equation}
\begin{split}
    \rho    =   a_0 |g\rangle \langle g| &+ \sum_{j} v_j^* |g\rangle \langle e_j| + \sum_{\mu} w_{\mu}^* |g\rangle \langle ee_{\mu}|\\
    + \sum_{j} v_j |e_j\rangle \langle g| &+ \sum_{i,j} \tilde{\rho}_{ij} |e_i\rangle \langle e_j| + \sum_{j,\mu} s_{j \mu}^* |e_j\rangle \langle ee_{\mu}| \\
    + \sum_{\mu} w_{\mu} |ee_{\mu}\rangle \langle g| &+ \sum_{j,\mu} s_{j \mu} |ee_{\mu}\rangle \langle e_j| + \sum_{\mu,\nu} \tilde{\tilde{\rho}}_{\mu\nu} |ee_{\mu}\rangle \langle ee_{\nu}|.\\
    \label{eqn:rho}
\end{split}
\end{equation}

The dynamics of the system will be calculated by using the Hamiltonian and the Lindblad superoperator
\begin{equation}
    \frac{d\hat{\rho}}{dt} = -\frac{i}{\hbar}[\hat{H}_{eff}, \hat{\rho}] + \mathcal{L} (\hat{\rho}),
    \label{eqn:rhodot}
\end{equation}
where $\hat{H}_{eff} = \hat{H}_{L} + \hat{H}_{dd}$, which are defined in Eqs. \eqref{eqn:HL} and \eqref{eqn:Hdd}, and $\mathcal{L}(\hat{\rho})$ is defined in Eq. \eqref{eqn:Lindblad}

The laser interaction is described by the Laser Hamiltonian in the rotating wave approximation given by 
\begin{equation}
    \hat{H}_L =  \hbar \sum_j \left[ -\delta\hat{\sigma}_j^+\hat{\sigma}_j^- + \frac{\Omega_j}{2}\hat{\sigma}_j^+ + \frac{\Omega^*_j}{2}\hat{\sigma}_j^- \right],
    \label{eqn:HL}
\end{equation}
where $\Omega_j = \Omega_0 e^{i \mathbf{k}_0\cdot \mathbf{r}_j}$ is the Rabi frequency of atom $j$, which governs the intensity of the driving laser and $\delta$ is the detuning. $\mathbf{r_j}$ gives the position of the $j$th atom. The laser Hamiltonian can induce transitions from the ground state to the first excited state and also from the first excited state to the second excited state.

Since this study is restricted to the low-intensity limit, the Rabi frequency should be small: $\Omega_0 \ll \Gamma_0$, where $\Gamma_0$ is the decay rate of a single atom. Even beyond this limit, the Rabi frequency should be smaller than the eigenmode decay rates: $\Omega_0 \ll \gamma_\alpha$ to ensure that there are only one or two photons in the case of long-lived eigenmodes (see Sec. \ref{sec:eigenmodes}).
The probability of excitation in the first excited state is proportional to $\Omega_0^2$, while the probability of excitation in the second excited state is of order $\Omega_0^4$.

In the semiclassical limit, the collective interactions between the atoms can be obtained by tracing out the vacuum radiation fields \cite{r2023,l1970}.
The coupling between atoms can then be described by the dyadic Green's function in vacuum $g(\mathbf{r}_{ij})$ given by,
\begin{equation}
\begin{split}
    g(\mathbf{r}_{ij}) = \frac{-i\Gamma_0}{2}&\bigg[h_0^{(1)}( kr_{ij}) + 
     \frac{3(\hat{r}_{ij}\cdot\hat{q})(\hat{r}_{ij}\cdot\hat{q}^*) - 1}{2} h_2^{(1)}(kr_{ij})
     \bigg],
     \label{eqn:greens}
\end{split}
\end{equation}
where $\hat{q}$ is the dipole orientation of the atoms, $\mathbf{r}_{ij} = \mathbf{r}_i - \mathbf{r}_j$ denotes the distance between atoms $i$ and $j$, $r_{ij} = |\mathbf{r}_{ij} |$ is the norm of $\mathbf{r}_{ij}$, $\hat{r}_{ij} = \mathbf{r}_{ij}/r_{ij}$ is the unit vector along $\mathbf{r}_{ij}$. $\Gamma_0$ is the decay rate of a single atom and $h_l^{(1)}$ are the outgoing spherical Hankel function of angular momentum $l$. 
The Green's function $g(\mathbf{r}_{ij})$ is proportional to the electric field of a dipole at position $\mathbf{r}_j$ due to another dipole at $\mathbf{r}_i$ \cite{jackson}.

When $i = j$, that is when $\mathbf{r}_{ij} = 0$, the real part of the function becomes undefined, while the imaginary part remains finite (${Im\{g(0)\} = -\Gamma_0/2}$). Hence, we redefine $g(\mathbf{r}_{ij})$ to be
\begin{equation}
\begin{aligned}
    g(\mathbf{r}_{ij}) &=  g(\mathbf{r}_{ij})&&for \quad i \neq j \\
    &=-\frac{i\Gamma_0}{2} &&for \quad  i = j.
\end{aligned}
\end{equation}

The real part of the Green's function captures the coherent dipole–dipole interactions corresponding to excitation exchange between atoms through virtual photons. This process leads to collective energy shifts of the atomic resonance \cite{fhm1973,s2009,mss2014}. This is described by the coherent dipole-dipole coupling Hamiltonian, 
\begin{equation}
    \hat{H}_{dd} = \hbar \sum_{i \neq j} Re\{g(\mathbf{r}_{ij}) \}\hat{\sigma}_i^+ \hat{\sigma}_j^-.
    \label{eqn:Hdd}
\end{equation}

The imaginary part of the Green's function, on the other hand, describes the dynamics of the de-excitation in the system and is modeled by the Lindblad operator given by
\begin{equation}
\begin{split}
    \mathcal{L}(\hat{\rho}) = \sum_{i,j} -Im\{g(\mathbf{r}_{ij})\}\big[&
    2\hat{\sigma}_i^- \hat{\rho} \hat{\sigma}_j^+  
    - \hat{\sigma}_j^+ \hat{\sigma}_i^- \hat{\rho} 
    - \hat{\rho}\hat{\sigma}_j^+ \hat{\sigma}_i^-
    \big],
    \label{eqn:Lindblad}
\end{split}
\end{equation}
where the $i = j$ terms describe the single atom spontaneous emission and the $i \neq j$ terms describe the collective spontaneous emission responsible for modifying the decay rates, resulting in subradiance or superradiance. The first term ($\hat{\sigma}_i^- \hat{\rho} \hat{\sigma}_j^+$) in particular is responsible for inducing a de-excitation of the atom, which is accompanied by the emission of a photon.

\subsection{Eigenmodes\label{sec:eigenmodes}}
The Green's function of the interaction between the atoms can be used to construct a matrix $G_{ij} = g(\mathbf{r}_{ij})$ that, when diagonalized, will give the natural single-excitation eigenmodes of the system. Each eigenvector $\mathbf{V}_{ \alpha}$ corresponds to a mode of the system with a particular decay rate given by the imaginary part of the eigenvalue $\gamma_{\alpha}$ and the energy shift given by the real part $\Delta_\alpha$.
\begin{equation}
    \sum_{j} G_{ij} \mathbf{V}_{j\alpha} = \mathcal{G}_\alpha \mathbf{V}_{i\alpha} = ( \Delta_{\alpha} - i\frac{\gamma_{\alpha}}{2})\mathbf{V}_{i\alpha}.
\end{equation}
The index $\alpha$ will be used in general to denote the single-excitation eigenmodes.
Since the $G_{ij}$ matrix is complex symmetric, the orthogonality conditions are different from the conventional eigenmodes of Hermitian matrices.
The eigenmodes can be normalized to follow 
\begin{equation}
    \sum_{i} \mathbf{V}_{i\alpha} \mathbf{V}_{i\alpha'} = \delta_{\alpha, \alpha'}.
    \label{eqn:orthogonality}
\end{equation}


Similar to how the single-excitation Green’s function describes the transfer of an excitation between two atoms, the double-excitation Green’s function characterizes this process in the presence of an additional excitation on an unrelated atom. It employs the same Green’s function expression [Eq. \eqref{eqn:greens}] to connect pairs of double-excitation states that share one common excited atom. In other words, for $\mu = (m_1,m_2)$ and $\nu = (n_1,n_2)$, a transition from $|ee_{\mu}\rangle$ to $|ee_{\nu}\rangle$  occurs through a single photon-exchange process described by the jump operator $\hat{\sigma}_{m_1}^+\hat{\sigma}_{n_1}^-$ if $ m_2 = n_2$.
Since the two-excitation states are twice as likely to decay, the diagonal terms will be $2 \times Im\{g(0)\} = -\Gamma_0$.
\begin{equation}
\begin{aligned}
    \tilde{G}_{\mu\nu} &=  g(\mathbf{r}_{m_1n_1})&&for \quad m_2 = n_2, m_1 \neq n_1 \\
    &=  g(\mathbf{r}_{m_2n_2}) &&for \quad m_1 = n_1, m_2 \neq n_2 \\
    &=  g(\mathbf{r}_{m_2n_1}) &&for \quad m_1 = n_2, m_2 \neq n_1 \\
    &=  g(\mathbf{r}_{m_1n_2}) &&for \quad m_2 = n_1, m_1 \neq n_2 \\
    &=-i\Gamma_0 &&for \quad m_1 = n_1, m_2 = n_2 \\
    &= 0 &&for \quad m_1 \neq n_1, m_2 \neq n_2. \\
\end{aligned}
\end{equation}
This can be diagonalized to obtain the eigenmodes that correspond to the doubly-excited states.
The index $\beta$ will be used to generally denote two-excitation eigenmodes. These eigenvectors $\mathbf{W}_{\mu \beta}$ also follow a similar situation where the imaginary part of the eigenvalue denotes the decay rate of a single photon being emitted and descending into single-excitation states.
\begin{equation}
    \sum_{\nu} \tilde{G}_{\mu\nu} \mathbf{W}_{\nu\beta} = \mathcal{G}_\beta^{(2)} \mathbf{W}_{\mu\beta} =  ( \Delta_{\beta}^{(2)} -i\frac{\gamma_{\beta}^{(2)}}{2})\mathbf{W}_{\mu\beta}.
\end{equation}
It is important to note that the $\gamma_{\beta}^{(2)}$ describes the rate at which the first photon is emitted. Each such state will then connect to a mix of single-excitation states after the emission. This diagonalization step is the most computationally intensive, scaling as $\mathcal{O}(N^6)$, and limits the maximum number of atoms that can be simulated to around 30.

\subsection{Calculation of $g^{(2)}$ correlation}
To calculate the $g^{(2)}$, the system is driven with a laser until it reaches steady state up to time $t$. The density matrix is then projected into a state where the system has emitted a photon \cite{loudon,reset-matrix,zmw1987,csv1989}. 
\begin{equation}
    \frac{\hat{\sigma}^- \hat{\rho}(t) \hat{\sigma}^+}{\langle\hat{\sigma}^+ \hat{\sigma}^-\rangle(t)} \rightarrow \hat{\rho}'(t),
    \label{eqn:ProjectedDM}
\end{equation}
where $\hat{\sigma}^-$ can represent the emission of a photon in a particular direction $\mathbf{k}$ or from a particular eigenmode $\alpha$,
\begin{equation}
    \hat{\sigma}^-_{\mathbf{k}} = \sum_{j} e^{-i \mathbf{k}. \mathbf{r}_j} \hat{\sigma}^-_j \quad OR \quad \hat{\sigma}^-_{\alpha} = \sum_{j} V_{j \alpha} \hat{\sigma}^-_j.
\end{equation}
Projecting the state due to emission from a specific eigenmode allows us to isolate how the system’s lifetime influences the photon correlations.

This projected density matrix $\hat{\rho}'$ is then evolved again in time using Eq. \eqref{eqn:rhodot} up to time $t + \tau$. Then the probability of another photon being emitted is calculated and normalized with the intensity

\begin{equation}
    g^{(2)}(\tau) = \frac{Tr[\hat{\sigma}^- \hat{\rho}'(t+\tau) \hat{\sigma}^+]}{\langle\hat{\sigma}^+ \hat{\sigma}^-\rangle(t)},
    \label{eqn:g2_1}
\end{equation}
which, by expanding the projected density matrix $\hat{\rho}'$ according to Eq. \eqref{eqn:ProjectedDM}, can also be written as
\begin{equation}
    g^{(2)}(\tau) = \frac{Tr[\hat{\sigma}^-(t+\tau)\hat{\sigma}^-(t) \hat{\rho} \hat{\sigma}^+(t)\hat{\sigma}^+(t+\tau)]}{[\langle\hat{\sigma}^+ \hat{\sigma}^-\rangle(t)]^2}.
    \label{eqn:g2_2}
\end{equation}

In many cases, we focus on the $g^{(2)}$ when $\tau = 0$, that is without any time delay, or the instantaneous two-photon correlation. That simplifies Eq. \eqref{eqn:g2_2} as
\begin{equation}
    g^{(2)}(\tau = 0) = \frac{Tr[\hat{\sigma}^-\hat{\sigma}^- \hat{\rho} \hat{\sigma}^+\hat{\sigma}^+]}{\langle\hat{\sigma}^+ \hat{\sigma}^-\rangle^2}.
    \label{eqn:g2_3}
\end{equation}

If the system is excited to a particular eigenmode $\alpha$ using the operator $\hat{\sigma}^+_{\alpha} = \sum_{j} V_{j \alpha} \hat{\sigma}^+_j$, we cannot use the same set of $\hat{\sigma}^+_{\alpha}$, $\hat{\sigma}^-_{\alpha}$ for the detection of light emitted in that mode. Due to the unusual orthogonality relation defined in Eq. \eqref{eqn:orthogonality}, the raising and lowering operators used for the detection have to be redefined as $\hat{\sigma}'^+_{\alpha} = \sum_{j} V^*_{j \alpha} \hat{\sigma}^+_j$, $\hat{\sigma}'^-_{\alpha} = (\hat{\sigma}'^+_{\alpha})^*$ to give 
$\hat{\sigma}'^-_{\alpha'}\hat{\sigma}^+_{\alpha}|g\rangle = \delta_{\alpha,\alpha'}|g\rangle$.

The results from the double excitation model for low intensities has been tested for convergence with a full density matrix treatment with small atom numbers.

\section{Results}

Using the formalism described in Sec. \ref{sec:methods}, the density matrix can be evolved to study the dynamics of both the single and the double-excitation states.
This paper will focus exclusively on correlations in the emitted light, and the correlations with the driving light fields will be ignored.


To keep the data shown consistent, the configuration of the atoms will be kept the same in this section. There are $25$ atoms in a square lattice in the XZ plane with lattice separation $d$ ranging from $ 0.3 \lambda$ to $1.0 \lambda$. The minimum separation is limited to $ 0.3 \lambda$, since smaller separations can cause the coherent dipole-dipole coupling to be larger than $\Gamma_0$, resulting in other effects like the dipole blockade \cite{wb2020}.
The atoms are polarized along the Z-axis. Although the data shown are only for a two-dimensional array, the results are qualitatively similar for other atom configurations.

When collective dipole interactions are absent, the steady-state density matrix contains no inter-atomic coherences ($\langle\hat{\sigma}_i^+\hat{\sigma}_j^- \rangle = 0$ for $i \neq j$), corresponding to independent emitters with only incoherent emission. In this limit, the $g^{(2)}(0)$ will depend on the number of atoms as $(1-1/N)$ \footnote{This $1-1/N$ expression corresponds to independent emitters in the weak-driving steady-state regime with no inter-atomic coherences. Independent emitters can exhibit a different zero-delay correlation of $2(1-1/N)$ in regimes of strong driving or when emission into a common indistinguishable mode produces interference \cite{agp2011}.}. 
The results will be compared to this independent-emitter reference value, which equals $0.96$ for $N = 25$ corresponding to nearly Poissonian photon statistics.



\subsection{Double-excitation eigenmodes\label{sec:DoubleEx}}


In this section, we will provide some basic descriptions of the second excitation eigenmodes. We will also explore the connection between the single and doubly excited states. A detailed investigation of the characteristics of the double-excitation eigenmodes has been done in \cite{poddubny1_2019}. Even though their description is for Waveguide-qubit systems, many of the properties are similar to atoms in free space. 


The most subradiant double-excitation state decays faster than the most subradiant single-excitation state in a single configuration of atoms in space. 
When two atoms are being excited, each excitation can only hop to the other N-2 atoms that are not excited, which is analogous to the first excited atom being invisible to the second excited atom. This 'hole' interferes with the perfect destructive interference necessary to get to the most subradiant state. The alternate is true for the superradiant state where the most superradiant double-excitation state decays faster than the most superradiant single-excitation state. 

The manifold of the double-excitation states is of $N(N-1)/2$ dimensions and the corresponding eigenmodes can be renormalized to form a complete ortho-normal subspace with the new definition of the inner product as in Eq. \eqref{eqn:orthogonality}. On the other hand, there are $N^2/2$ vectors that can be made when any two single-excitation eigenmodes are used to excite the ground state procedurally. This discrepancy in the number of states of the double-excitation eigenmodes and two single excitations comes about because there cannot be two excitations on a single atom, i.e., $\hat{\sigma}_i^+\hat{\sigma}_i^+|g\rangle = 0$.

\begin{figure}
    \centering
    \includegraphics[width=1.0\linewidth]{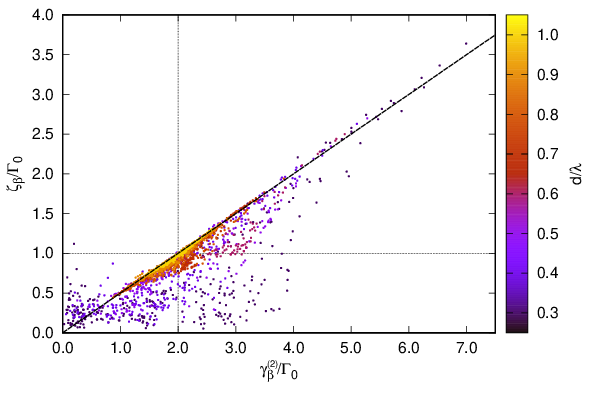}
    \caption{The decay rate of the second photon ($\zeta_\beta$) versus the decay rate of the first photon ($\gamma_{\beta}^{(2)}$) from a double-excitation eigenmode $\beta$. The separation has been varied from 0.3 to 1.0 $\lambda$. The color shows the separation $d$ of the atoms in the array. The dashed line corresponds to $\zeta_\beta = \gamma_{\beta}^{(2)}/2$. }
    \label{fig:Zeta}
\end{figure}

The imaginary part of the eigenmode $\gamma_{\beta}^{(2)}$ gives the rate of decay of the double-excitation eigenmode. Once the first photon is emitted into free space, the system is brought into a combination of single-excitation states. This combination of single-excitation states will then result in a corresponding average decay rate for the second photon. 
This decay rate of the second photon will be represented by $\zeta_\beta$ has been plotted in Fig. \ref{fig:Zeta} and is defined by,
\begin{equation}
    \zeta_\beta = \frac{4}{{\gamma_{\beta}^{(2)}}}Tr[\sum_{iji'j'} Im\{g(\mathbf{r}_{i'j'})\}  Im\{g(\mathbf{r}_{ij})\}  \hat{\sigma}_{i'}^-\hat{\sigma}_{i}^- \rho_\beta \hat{\sigma}_{j}^+\hat{\sigma}_{j'}^+],
\end{equation}
where $\rho_\beta$ is the density matrix initialized to contain only the second excitation eigenmode $\beta$.
In most cases, the $\zeta_\beta$ is approximately close to $\gamma_{\beta}^{(2)}/2$, especially for higher atom separations. There are exceptions to this when the atom separations get smaller. Specifically, for states that have very small $\gamma_{\beta}^{(2)}$, the $\zeta_\beta$'s are not as suppressed and tend to go above the line that depicts $\zeta_\beta = \gamma_{\beta}^{(2)}/2$ in Fig. \ref{fig:Zeta}. This becomes more relevant in Sec. \ref{sec:DetectFS}.

The single and double-excitation eigenmodes are vectors with different sizes and normalization schemes, which makes it hard to draw conclusions on how they correlate with each other. 
To understand this, we can define a quantity that describes the overlap of the double-excitation eigenmode $\beta$ with two single-excitation eigenmodes $\alpha_1$ and $\alpha_2$. This can also be understood as the tendency of mode $\beta$ to emit two photons, one in mode $\alpha_1$ and one in mode $\alpha_2$. 
\begin{equation}
    L_{\alpha_1 \alpha_2 \beta} = \sum_{\mu} (V_{m_{1}\alpha_1} V_{m_{2}\alpha_2} + V_{m_{1}\alpha_2} V_{m_{2}\alpha_1}) W_{\mu\beta},
    \label{eqn:Laab}
\end{equation}
where the index $\mu = (m_1,m_2)$ represents the atoms $m_1$ and $m_2$ being excited [Refer Fig. \ref{fig:schematic}].

When $\alpha_1 = \alpha_2$, it can be redefined as $X_{\alpha\beta}$ which is analogous to a projection of $(V_{i\alpha})^2$ over $W_{\mu\beta}$. This describes the geometric overlap of the double-excitation eigenmode with the state where both photons are in the same single-excitation eigenmode. This quantity will become relevant in Sec. \ref{sec:SingleMode}.
\begin{equation}
    X_{\alpha\beta} = \sum_{\mu} (2 V_{m_{1}\alpha} V_{m_{2}\alpha}) W_{\mu\beta}.
    \label{eqn:Xab}
\end{equation}

\begin{figure}
    \centering
    \includegraphics[width=0.99\linewidth]{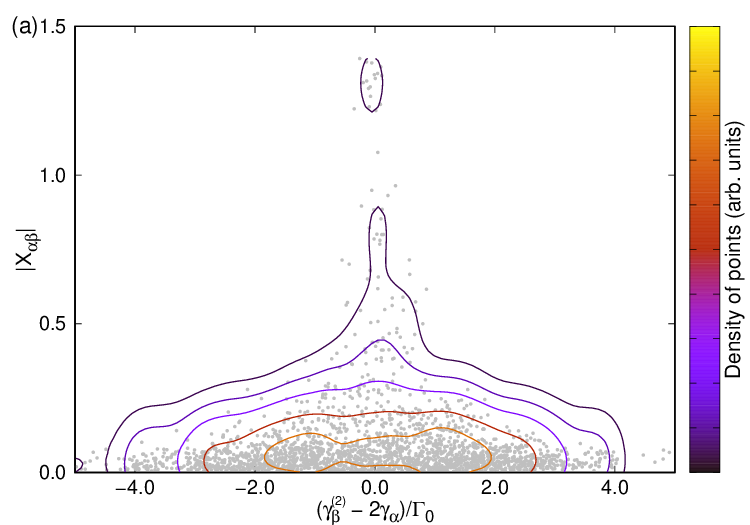}
    \includegraphics[width=0.99\linewidth]{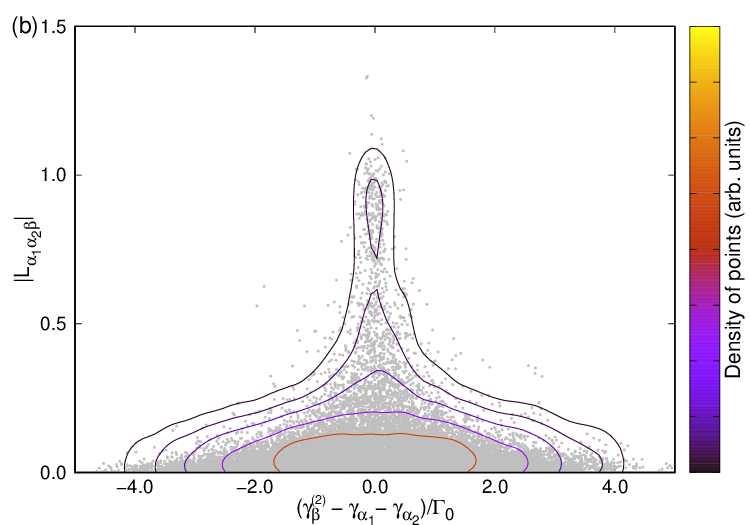}
    \caption{The overlap between single and double-excitation eigenmodes. (a) The $\vert X_{\alpha\beta}\vert $ of the single-excitation eigenmode $\alpha$ and double-excitation eigenmode $\beta$ versus the difference in decay rate  $(\gamma_{\beta}^{(2)} - 2 \gamma_\alpha)/\Gamma_0$.
    (b) The $\vert L_{\alpha_1 \alpha_2 \beta}\vert $ of the single-excitation eigenmodes $\alpha_1$, $\alpha_2$ and the double-excitation eigenmode $\beta$ versus the difference in decay rate  $(\gamma_{\beta}^{(2)} -  \gamma_{\alpha_1} - \gamma_{\alpha_2})/\Gamma_0$, where $\alpha_1 \neq \alpha_2$. The data is shown for inter-atom separation of $d = 0.4 \lambda$. The contour depicts the density of points. The data is calculated for a square array of 25 atoms.}
    \label{fig:Xab}
\end{figure}

Figure \ref{fig:Xab} shows the $\vert X_{\alpha\beta}\vert $ and $\vert L_{\alpha_1 \alpha_2 \beta}\vert $ ($\alpha_1 \neq \alpha_2$) for a square array with 25 atoms with separation of $0.4\lambda$. The points correspond to the N single-excitation eigenmodes $\alpha$ and $N(N-1)/2$ double-excitation eigenmodes $\beta$.

Although the overlaps $\vert X_{\alpha\beta}\vert $ and $\vert L_{\alpha_1 \alpha_2 \beta}\vert $ only depend on the spatial profile of the eigenmodes, they are inherently still connected to the decay rates of the eigenmodes. In both plots, the overlap is maximized when the sum of the single-excitation decay rates matches the double-excitation decay rates. In the case of$\vert X_{\alpha\beta}\vert $, the overlap is maximum when $\gamma_{\beta}^{(2)} = 2\gamma_\alpha$.

The spread of the points along the x-axis (difference in decay rate) depends on the separation of the atoms $d$. 
As the separation becomes smaller, the range of decay rates increases, i.e., better subradiant and superradiant states are achievable. This makes the spread of the difference in decay rate also larger. This means that non-negligible overlap is possible with larger differences in decay rates for smaller separations.

Even with small differences in decay rates, many points exhibit only a small overlap. This suggests that a small decay rate difference does not guarantee good overlap, although good overlap tends to occur when decay rate differences are small.

\subsection{Single mode excitation\label{sec:SingleMode}}


In experiments, it is usually possible to detect/input light from only a single mode (a common example is a Gaussian light mode). But using the geometry of the atom ensemble, it can be made to excite eigenmodes with different decay rates. For example, when using Gaussian light to excite and detect light from a finite array of atoms, having the separation between the atoms to be less than half a wavelength will result in the superradiant modes being predominantly excited. The opposite is true for separations in the range of half to one wavelength where the subradiant modes are excited \cite{bloch}. 

Hence we first consider a situation where we are exciting and detecting the same mode of light, but each mode will be an eigenmode of the atomic ensemble. While most of the eigenmodes of atomic systems in free space, especially the subradiant modees, will be difficult to directly excite in experiments, we are studying this to isolate the effect of only the decay rate on the correlations in the emitted light. There are also platforms such as Transmon Waveguide-QED, which offer a high degree of individual emitter control while still supporting strong collective effects, allowing direct access to specific eigenmodes.

In the calculations, hypothetical lasers that can exactly address a particular single-excitation eigenmode of the system are used. The corresponding spatial profile and detuning of the eigenmode can be imprinted on the driving laser. 
After emitting a photon, the time the system takes to recover back to steady state is usually determined by the eigenmode's decay rate.

On the other hand, the zero-time-delay photon correlation $g^{(2)}(0)$ does not trivially depend on this decay rate. Since $g^{(2)}(0)$ describes how likely it is to emit two photons together, the decay rates of the double-excitation eigenmodes will also contribute. Therefore, we numerically and analytically study the dependence of the zero-time-delay second-order photon correlation on the decay rates of the single and double-excitation eigenmodes. 

\subsubsection{Detection in the same Eigenmode $\alpha$\label{sec:DetectAlpha}}

In this section, the $g^{(2)}(0)$ will be calculated when the light is emitted into the same single-excitation eigenmode with which it is excited. This represents how two photons each in a single-excitation eigenmode $V_{i\alpha}$ connects with the double-excitation modes $W_{\mu\beta}$, which was described by $X_{\alpha\beta}$ as discussed in Sec. \ref{sec:DoubleEx} and in Fig. \ref{fig:Xab}(a).


While the $g^{(2)}(0)$ can be calculated numerically by time evolving the density matrix to steady state and taking projections, we can simplify the calculation by using eigenmode decomposition. An analytical expression can be derived for the steady state and $g^{(2)}(0)$ when exciting using a single eigenmode. The derivation is detailed in the Appendix \ref{app:singlemode}.



The $g^{(2)}(0)$ when driven and detected using the single-excitation eigenmode $\alpha$ with decay rate $\gamma_{\alpha}$ and a detuning corresponding its line shift $\delta = \Delta_{\alpha}$ is given by,
\begin{equation}
    g^{(2)}(0) = \gamma_{\alpha}^2 \bigg | \sum_{\beta} \frac{X_{\alpha \beta}^2}{\gamma_{\beta}^{(2)} + 2i(\Delta_{\beta}^{(2)} - 2\Delta_\alpha)} \bigg |^2,
    \label{eqn:g20_single}
\end{equation}
where the sum over $\beta$ goes over all the double-excitation eigenmodes. 

\begin{figure}
    \centering
    \includegraphics[width=1.0\linewidth]{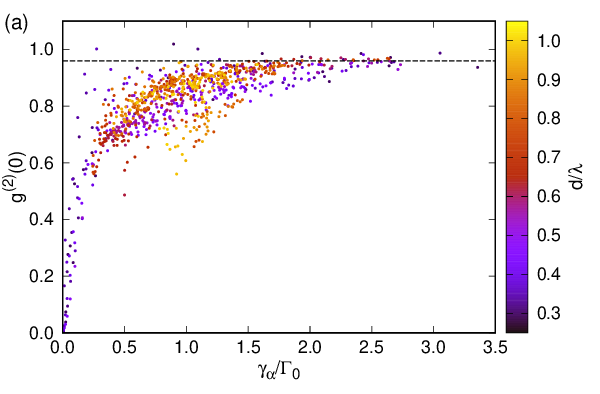}
    \includegraphics[width=1.0\linewidth]{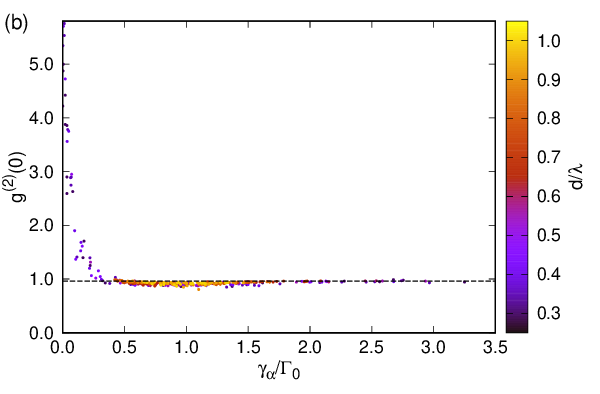}
    \caption{The $g^{(2)}(0)$ when excited using a single eigenmode $\alpha$ for an ensemble of 25 atoms arranged in a square array, versus the decay rate of the eigenmode ($\gamma_\alpha$). (a) depicts the situation when the detection is also in mode $\alpha$, where subradiant states exhibit anti-bunching despite the presence of double-excitation states. (b) corresponds to detection over all of free space, in which subradiant states instead display bunched emission. The separation has been varied from 0.3 to 1.0 $\lambda$. The color shows the separation $d$ of the atoms in the array. The dotted line shows the independent-emitter incoherent emission $g^{(2)}(0)$ for $25$ atoms.}
    \label{fig:g2SingleMode}
\end{figure}

The $g^{(2)}(0)$ can be understood as the ratio between the decay rates of the single-excitation eigenmodes over the decay rates of the double-excitation eigenmodes weighted over the coupling coefficient $X_{\alpha\beta}$. The detuning also plays a role in decreasing how well the double-excitation eigenmode is excited. 
As seen in Fig. \ref{fig:Xab}(a), the coupling coefficient $X_{\alpha\beta}$ only has a significant contribution when $\gamma_{\alpha} \approx \gamma_{\beta}^{(2)}/2$ resulting in the $g^{(2)}(0)$ almost always being less than 1.


For highly subradiant states, the $g^{(2)}(0)$ is very low and is almost proportional to the decay rate. This shows a clear correlation between antibunching and low-intensity subradiance. But once the decay rate reaches $0.5\Gamma_0$, the $g^{(2)}(0)$ starts to saturate and approaches the independent-emitter incoherent emission value of $0.96$. Even for highly superradiant states, the $g^{(2)}(0)$ is only around or less than this value. This means that, unlike in the high-intensity situation where superradiance implies bunching, superradiance in the low-intensity limit only makes the photon emission Poissonian. 

This effect is a consequence of the system being at steady state, where the dependence on the decay rates plays a role. When the system is excited using very short pulses (where the decay rate is irrelevant in determining the excitation of the system), both the subradiant and superradiant states emit close to Poissonian light.


\subsubsection{Detection in all of Free space\label{sec:DetectFS}}

So far, we have described the photons being detected in the same mode they were excited in. 
In this section, the detection scheme covers the light emitted into all of free space.
This type of detection is more relevant in some experimental implementations, especially in Waveguide QED where all the light being emitted can be easily detected from the two ends of the waveguide. In this scenario, the emission of both photons is not restricted to the same mode $\alpha_1$ but includes the possibility of emission into a combination of the eigenmodes $\alpha_1$ and $\alpha_2$.


The total intensity of light can be calculated from the surrounding term of the Lindblad operator $2Re\{g(\mathbf{r}_{ij})\}\hat{\sigma}_i^- \hat{\rho} \hat{\sigma}_j^+$ that represents the emission of light and lowering of the excitation in the system. Since the eigenmodes of $g(\mathbf{r}_{ij})$ discussed so far will no longer be eigenmodes of this decay operator $2Re\{g(\mathbf{r}_{ij})\}$, we cannot derive a simple relation like Eq. \eqref{eqn:g20_single}. Nevertheless, it can be directly calculated and we can draw simple conclusions.

The results have been plotted in Fig. \ref{fig:g2SingleMode} (b) for a similar configuration. Contrary to Fig. \ref{fig:g2SingleMode} (a), the $g^{(2)}(0)$ in this case starts as a bunched emission for subradiant states and approaches $0.96$ again for superradiant states. 
The initial time emission correlation $g^{(2)}(0)$ describes the possibility of the second photon being emitted immediately after the first photon is emitted.
As seen in Fig. \ref{fig:Zeta}, for a double-excitation eigenmode $\beta$, the second photon decay rate $\zeta_\beta$ follows the line $\zeta_\beta = \gamma_{\beta}^{(2)} / 2$ for large $\gamma_{\beta}^{(2)}$ but fails to keep up as it reaches very small decay rates.
When excited using subradiant states, even though the decay rate of the first photon is small, the second photon rate is comparatively larger, leading to larger $g^{(2)}(0)$ and bunched emission.

Unlike the previous section, this effect is not dependent on reaching steady state and will occur even when driven using short pulses.

This result contrasts with Ref. \cite{wb2020}, which reports antibunching in subradiant states at very small atomic separations. That observation pertains to a different regime, known as the \textit{dipole blockade}, which arises from large interaction-induced energy shifts when atoms are closely spaced ($d < 0.2 \lambda$). This effect can be understood from the denominators in Eqs. \ref{eqn:g20_single} and \ref{eqn:app:EigContib}, where significant detuning mismatches suppress double-excitation probabilities, thereby inducing a blockade.

In conclusion, when exciting the system using a single characteristic eigenmode $\alpha$, the trend of the $g^{(2)}(0)$ versus the decay rate $\gamma_\alpha$ depends on the type of detection of the photons. For subradiant states, detection in the same eigenmode $\alpha$ displays anti-bunched emission, while detection in all free spaces results in bunched emission. However, superradiant states tend to converge to the independent-emitter incoherent emission $g^{(2)}(0) = 1-1/N$ irrespective of the detection scheme. The behavior of $g^{(2)}(0)$ as N increases is governed by the scaling of the corresponding subradiant lifetimes. In general, larger ensembles support enhanced subradiance, but the extent of this enhancement is strongly dependent on the geometry of the array.

\subsection{Two interfering modes\label{sec:DoubleMode}}

In this section, we will discuss how two different single-excitation eigenmodes interfere and interact with the double-excitation eigenmodes. 
In Sec. \ref{sec:DetectAlpha}, the system was both excited and detected using the same eigenmode $\alpha$. Another eigenmode $\tilde{\alpha}$ will be added to the excitation with a relative phase, and the dependence of the $g^{(2)}(0)$ will be studied.
The Rabi frequency for each atom will be given by,
\begin{equation}
    \Omega_j = \Omega_0(\mathbf{V}_{j{\alpha}} + A e^{i\phi} \mathbf{V}_{j{\tilde{\alpha}}}),
\end{equation}
where $\phi$ is the relative phase, and $A$ is relative amplitude of the mode $\tilde{\alpha}$.

When the two eigenmodes interfere, the orthogonality ensures that the populations in the single-excitation eigenmodes do not vary with the relative phase $\phi$. However, in the case of the double-excitation eigenmodes, the correlations are emphasized and affect their population. 
Since the different double-excitation modes couple differently to the original mode of detection, $g^{(2)}(0)$ can vary without a change in intensity. 

\begin{figure}
    \centering
    \includegraphics[width=0.98\linewidth]{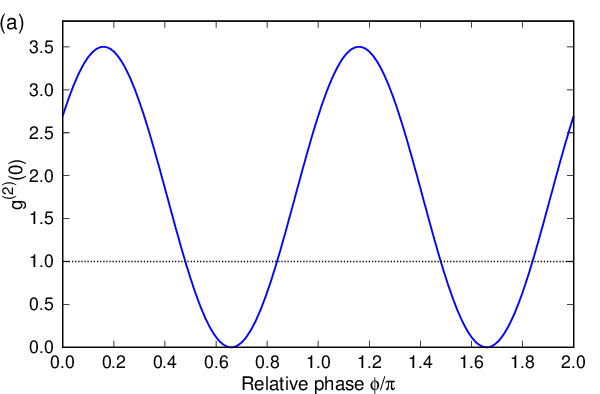}
    \includegraphics[width=0.98\linewidth]{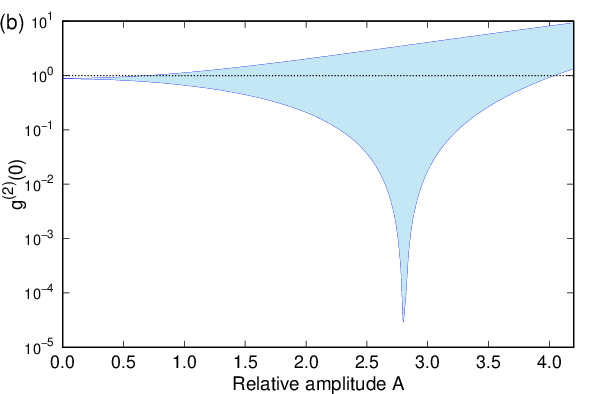}
    \caption{The $g^{(2)}(0)$ when two eigenmodes are incident and the light emitted into one mode is detected. (a) depicts the dependence on the relative phase $\phi$ at relative amplitude $A = 2.8$ and (b) depicts the maximum and minimum range of $g^{(2)}(0)$ possible when the relative amplitude $A$ is varied. The black dotted line is a reference for $g^{(2)}(0) = 1$.}
    \label{fig:2Mode}
\end{figure}

For a system with 25 atoms in a square array with separation $d = 0.4 \lambda$ and relative amplitude $A = 2.8$, the $g^{(2)}(0)$ oscillates between $0.0002$ and $3.5$ when using the most superradiant and most subradiant eigenstates as the two interfering modes. This has been depicted in Fig. \ref{fig:2Mode}(a). The maximum and minimum $g^{(2)}(0)$ possible for different relative amplitudes is depicted in Fig. \ref{fig:2Mode}(b). 
The recovery time of $g^{(2)}(\tau)$ will simply be the decay rate of the mode that is being detected. With the relative phase as a knob, the $g^{(2)}(0)$ of the emitted light can be controlled. This can also be thought of as a controllable nonlinearity. 

The analytical form of $g^{(2)}(0)$ in this case can be derived similar to App. \ref{app:singlemode}. Since there are two eigenmodes, the expression $L_{\alpha \tilde{\alpha} \beta}$ will also play a role. 
\begin{equation}
\begin{split}
    g^{(2)}(0) = \gamma_{\alpha}^2 \bigg | \sum_{\beta} \frac{X_{\alpha \beta}}{\gamma_{\beta}^{(2)} + 2i(\Delta_{\beta}^{(2)} - 2\Delta_\alpha)} \\
    \times \bigg( X_{\alpha \beta} + e^{2i\phi}\eta A^2 X_{\tilde{\alpha} \beta} + e^{i\phi} (1+\eta)A L_{\alpha \tilde{\alpha} \beta} \bigg )
    \bigg |^2,
    \label{eqn:g20_double}
\end{split}
\end{equation}
where $\alpha$ and $\tilde{\alpha}$ are the two eigenmodes used to excite the system and $\beta$ denotes the index of double-excitation. $A$ is the relative amplitude and $\eta$ is the ratio of decay rates of the two eigenmodes and will be given by $\eta = \gamma_\alpha / (\gamma_{\tilde{\alpha}} + 2i(\Delta_{\tilde{\alpha}} -\Delta_\alpha))$. 

This equation is exactly the same as Eq. \eqref{eqn:g20_single} when $A = 0$. When $A \neq 0$, only the second and third terms cause a dependence on $\phi$.
The three terms in the second line describe the contribution of the two photons from the excitation eigenmodes $\alpha$ and $\tilde{\alpha}$. 
The first term describes both photons being excited by mode $\alpha$. The second term describes both photons excited by mode $\tilde{\alpha}$ and depends on $A^2$ and $X_{\tilde{\alpha}\beta}$. The third term denotes the situation in which the two photons are each excited using $\alpha$ and $\tilde{\alpha}$. It depends on the parameter $L_{\alpha\tilde{\alpha}\beta}$ that was depicted in Eq. \eqref{eqn:Laab} and in Figs. \ref{fig:Xab}(b).

Although the ratio of decay rates $\eta$ contributes to the $\phi$ dependent terms, the choice of eigenmodes is flexible. Although the most superradiant and subradiant states were used in the example shown in Fig. \ref{fig:2Mode}, this effect can be observed with any combination of eigenmodes and large antibunching can be found at different relative amplitudes. Once the first eigenmode has been chosen, the second perturbing eigenmode can be any combination of all the other $N-1$ orthogonal eigenmodes. An optimum combination could be found to further suppress the $g^{(2)}(0)$ minima. 

Unlike in Sec. \ref{sec:DetectAlpha}, this effect does not depend on reaching steady state and can also be observed when the excitations are short pulses. This can be utilized in applications where single photon pulses are required. If the system is configured to be in a $g^{(2)}(0)$ minima, the two-photon emission probability can be highly suppressed while also allowing for an increase in the intensity of the incident light.

To get an intuition behind this phenomenon, we can study the simple case with two atoms. A similar concept has also been studied experimentally in Ref. \cite{wrv2020}.
The eigenmodes of the system are $|+\rangle = (|eg\rangle + |ge\rangle)/\sqrt{2}$ and $|-\rangle = (|eg\rangle - |ge\rangle)/\sqrt{2}$. By interfering two modes with an arbitrary relative phase, the minima of $\langle ee\rangle$ occurs when the modes cancel out on one atom and only the other atom interacts with the incident light. This means that the phase controls whether the incident light interacts with one or two atoms. While this seems simple, the situation becomes nontrivial when many atoms are involved, and the level of nonlinearity can be changed on a much larger scale. 

Similarly, when there are many atoms, the interference controls the populations in the different double-excitation eigenmodes. However, the caveat is that having a beam that can address or detect individual eigenmodes becomes difficult. In some cases, the variation in the spatial part of the eigenmode becomes so fast that the mode becomes dark and cannot be simply accessed with optical beams. Therefore, we need systems that are similar, yet have a greater degree of control than free space atoms. 



One such system with a higher degree of control is the Transmon Waveguide QED system. These are artificial atom-like systems made using superconducting circuits that can interact with each other through microwave waveguides. These transmons can also be directly driven using the electronics, in addition to the microwave driving through the waveguide. This facilitates individual addressing of the qubits that can be used to access any arbitrary eigenmode of the system.
This can then be the knob that can be used to control the $g^{(2)}$ emission of the waveguide mode. In situations where the transmons are coherently controlled using the waveguide photons, a second orthogonal mode could be driven directly using the electronics to suppress or enhance two-photon emissions.

\section{Conclusion}

We studied the photon statistics of the emitted light in collectively interacting dipole systems using double-excitation states. We explored how the decay rate of the system plays a role in determining the photon correlations in the emitted light.
We described the emission properties of the double-excitation eigenmodes and explored how they couple to the single-excitation eigenmodes.

When the system is excited by a single-excitation eigenmode, we study the $g^{(2)}(0)$ of the emission when the photons are detected in (i) the same excitation eigenmode or (ii) free space.
For superradiant states, regardless of the detection type, the emission is mostly Poissonian. For subradiant states, the emission is either (i) antibunched or (ii) bunched, depending on the type of detection. This is contrary to the many-photon situation where superradiance induces bunching and subradiance causes antibunching.

When the system is excited by two distinct single-excitation eigenmodes, the interference between the two can cause interesting dependencies on the photon statistics. By changing the phase between the two incident excitations, the nonlinearity of the system can be controlled to an extent, resulting in being able to arbitrarily control the $g^{(2)}(0)$ of the emitted light. This can be utilized to enhance or suppress two-photon emission and extend the intensity range for single-photon operations.


Most of the observed effects can be attributed to the coupling between the different sets of eigenmodes. This formalism can be easily extended to other types of atom-atom interactions once the corresponding eigenmodes and orthogonality relations can be established.

Data for the figures used in this publication is available at \cite{datalink}.

\begin{acknowledgments}
This work was supported by the National Science Foundation under Award No. 2410890-PHY. This research was supported in part through computational resources provided by Information Technology at Purdue University, West Lafayette, Indiana.

\end{acknowledgments}

\appendix

\section{Single mode emission analytical calculation\label{app:singlemode}}

The density matrix can be evolved in time using Eq. \eqref{eqn:rhodot} until it reaches steady state. However, this method becomes tedious to calculate when N increases or when dealing with highly subradiant modes. Hence we can use the low-intensity approximation and directly calculate the steady state using eigenstate decomposition. Each of the types of terms in the density matrix represented in Eq. \eqref{eqn:rho} can be transformed into the corresponding eigenmode basis.
\begin{equation}
\begin{split}
    \tilde{v}_\alpha = \sum_j v_j \mathbf{V}_{j\alpha}  \quad\quad
    v_j = \sum_\alpha \tilde{v}_\alpha \mathbf{V}_{j\alpha} \\
    \tilde{w}_\beta = \sum_\mu  w_\mu \mathbf{W}_{\mu\beta} \quad\quad
    w_\mu = \sum_\beta \tilde{w}_\beta \mathbf{W}_{\mu\beta} 
    \label{eqn:app:transform}
\end{split}
\end{equation}
and similar relations can be shown for the other terms.

In the low-intensity limit, the Rabi frequency is much smaller than the decay timescales. As a consequence, each term in Eq. \eqref{eqn:rho}, to the right or below is one order higher with respect to $\Omega$, i.e., $a_0 \propto \Omega^0$, $v_j \propto \Omega^1$, $w_{\mu}, \tilde{\rho}_{ij} \propto \Omega^2$, $s_{j\mu} \propto \Omega^3$ and $\tilde{\tilde{\rho}}_{\mu\nu} \propto \Omega^4$.
It can be shown that by ignoring the higher order $\Omega$ terms in the rate equations, the terms in Eq. \eqref{eqn:rho} become
\begin{equation}
    \tilde{\rho}_{ij} = \frac{v_i v^*_j}{a_0}\quad
    s_{j\mu} = \frac{w_\mu v^*_j}{a_0}\quad
    \tilde{\tilde{\rho}}_{\mu\nu} = \frac{w_\mu w^*_\nu}{a_0}
    \label{eqn:app:rhoterms}
\end{equation}
where $a_0$ is the population in the ground state, which can be calculated from $Tr[\rho] = 1$.

Using these Eqs. \eqref{eqn:rhodot}, \eqref{eqn:app:transform}, and \eqref{eqn:app:rhoterms}, the contribution from each eigenmode at steady state can be directly calculated from $\Omega_j$
\begin{equation}
\begin{split}
    &\tilde{v}_\alpha = \frac{a_0}{2} \frac{\sum_j \Omega_j \mathbf{V}_{j\alpha}}{(\mathcal{G}_\alpha - \delta)}\\
    &\tilde{w}_\beta = \frac{a_0}{2} \frac{\sum_\mu (\Omega_{m_1} v_{m_2} + \Omega_{m_2} v_{m_1})\mathbf{W}_{\mu\beta}}{(\mathcal{G}_\beta^{(2)} - 2\delta)}.
\end{split}
\end{equation}

When the system is excited using a particular single-excitation eigenmode $\tilde{\alpha}$, $\Omega_j = \Omega_0 \mathbf{V}_{j\tilde{\alpha}}$ and $\delta = \Delta_{\tilde{\alpha}}$ will give
\begin{equation}
\begin{split}
    &\tilde{v}_\alpha =  \frac{ia_0\Omega_0}{\gamma_{\tilde{\alpha}}}  \delta_{\alpha, \tilde{\alpha}} \\
    &\tilde{w}_\beta =  \frac{-a_0^2\Omega_0^2}{\gamma_{\tilde{\alpha}}} \frac{X_{\tilde{\alpha}\beta}}{\gamma_{\beta}^{(2)} + 2i(\Delta_{\beta}^{(2)} - 2\Delta_{\tilde{\alpha}})}.
    \label{eqn:app:EigContib}
\end{split}
\end{equation}

Using this, the coefficients of the final density matrix [Eq. \eqref{eqn:rho}] at steady state can be calculated. The coefficients $v_j$ and $w_\mu$ can be calculated using Eqs. \eqref{eqn:app:EigContib} and \eqref{eqn:app:transform}. The rest of the terms of the density matrix can be calculated using Eq. \eqref{eqn:app:rhoterms}.

For detecting the light emitted into the same eigenmode $\tilde{\alpha}$, we use $\hat{\sigma}^-_{\tilde{\alpha}} = \sum_{j} V_{j \tilde{\alpha}} \hat{\sigma}^-_j$ to arrive at Eq. \eqref{eqn:g20_single}

\begin{equation}
\begin{split}
    g^{(2)}(0) = &\frac{Tr\{\hat{\sigma}^-_{\tilde{\alpha}}\hat{\sigma}^-_{\tilde{\alpha}} \rho \hat{\sigma}^+_{\tilde{\alpha}}\hat{\sigma}^+_{\tilde{\alpha}} \}}{Tr\{\hat{\sigma}^-_{\tilde{\alpha}} \rho \hat{\sigma}^+_{\tilde{\alpha}} \}^2}\\ 
    = &\gamma_{\tilde{\alpha}}^2 \bigg | \sum_{\beta} \frac{X_{{\tilde{\alpha}} \beta}^2}{\gamma_{\beta}^{(2)} + 2i(\Delta_{\beta}^{(2)} - 2\Delta_{\tilde{\alpha}})} \bigg |^2.
    \label{eqn:appg2} 
\end{split}
\end{equation}

\bibliography{ref.bib}

@article{datalink,
  title = {Data for: Emission photon statistics in collectively interacting dipole atom arrays in the low-intensity limit},
  author = {Suresh, Deepak A. and Robicheaux, F.},
  journal = {}
}

@article{zoller,
  title = {Subradiant Bell States in Distant Atomic Arrays},
  author = {Guimond, P.-O. and Grankin, A. and Vasilyev, D. V. and Vermersch, B. and Zoller, P.},
  journal = {Phys. Rev. Lett.},
  volume = {122},
  issue = {9},
  pages = {093601},
  numpages = {6},
  year = {2019},
  month = {Mar},
  publisher = {American Physical Society},
  doi = {10.1103/PhysRevLett.122.093601},
  url = {https://link.aps.org/doi/10.1103/PhysRevLett.122.093601}
}

@Article{bloch,
author={Rui, Jun
and Wei, David
and Rubio-Abadal, Antonio
and Hollerith, Simon
and Zeiher, Johannes
and Stamper-Kurn, Dan M.
and Gross, Christian
and Bloch, Immanuel},
title={A subradiant optical mirror formed by a single structured atomic layer},
journal={Nature},
year={2020},
month={Jul},
day={01},
volume={583},
number={7816},
pages={369-374},
issn={1476-4687},
doi={10.1038/s41586-020-2463-x},
url={https://doi.org/10.1038/s41586-020-2463-x}
}

@article{bettles,
  title = {Cooperative ordering in lattices of interacting two-level dipoles},
  author = {Bettles, Robert J. and Gardiner, Simon A. and Adams, Charles S.},
  journal = {Phys. Rev. A},
  volume = {92},
  issue = {6},
  pages = {063822},
  numpages = {6},
  year = {2015},
  month = {Dec},
  publisher = {American Physical Society},
  doi = {10.1103/PhysRevA.92.063822},
  url = {https://link.aps.org/doi/10.1103/PhysRevA.92.063822}
}

@article{dicke,
  title = {Coherence in Spontaneous Radiation Processes},
  author = {Dicke, R. H.},
  journal = {Phys. Rev.},
  volume = {93},
  issue = {1},
  pages = {99--110},
  numpages = {0},
  year = {1954},
  month = {Jan},
  publisher = {American Physical Society},
  doi = {10.1103/PhysRev.93.99},
  url = {https://link.aps.org/doi/10.1103/PhysRev.93.99}
}

@article{ggv2018,
  title = {Free-space photonic quantum link and chiral quantum optics},
  author = {Grankin, A. and Guimond, P. O. and Vasilyev, D. V. and Vermersch, B. and Zoller, P.},
  journal = {Phys. Rev. A},
  volume = {98},
  issue = {4},
  pages = {043825},
  numpages = {17},
  year = {2018},
  month = {Oct},
  publisher = {American Physical Society},
  doi = {10.1103/PhysRevA.98.043825},
  url = {https://link.aps.org/doi/10.1103/PhysRevA.98.043825}
}

@article{re1971,
  title = {Superradiance},
  author = {Rehler, Nicholas E. and Eberly, Joseph H.},
  journal = {Phys. Rev. A},
  volume = {3},
  issue = {5},
  pages = {1735--1751},
  numpages = {0},
  year = {1971},
  month = {May},
  publisher = {American Physical Society},
  doi = {10.1103/PhysRevA.3.1735},
  url = {https://link.aps.org/doi/10.1103/PhysRevA.3.1735}
}

@article{fhm1973,
title = "Frequency shifts in emission and absorption by resonant systems ot two-level atoms",
journal = "Physics Reports",
volume = "7",
number = "3",
pages = "101 - 179",
year = "1973",
issn = "0370-1573",
doi = "https://doi.org/10.1016/0370-1573(73)90001-X",
url = "http://www.sciencedirect.com/science/article/pii/037015737390001X",
author = "R. Friedberg and S.R. Hartmann and J.T. Manassah"
}

@article{gfp1976,
  title = {Observation of Near-Infrared Dicke Superradiance on Cascading Transitions in Atomic Sodium},
  author = {Gross, M. and Fabre, C. and Pillet, P. and Haroche, S.},
  journal = {Phys. Rev. Lett.},
  volume = {36},
  issue = {17},
  pages = {1035--1038},
  numpages = {0},
  year = {1976},
  month = {Apr},
  publisher = {American Physical Society},
  doi = {10.1103/PhysRevLett.36.1035},
  url = {https://link.aps.org/doi/10.1103/PhysRevLett.36.1035}
}

@article{s2009,
  title = {Collective Lamb Shift in Single Photon Dicke Superradiance},
  author = {Scully, Marlan O.},
  journal = {Phys. Rev. Lett.},
  volume = {102},
  issue = {14},
  pages = {143601},
  numpages = {4},
  year = {2009},
  month = {Apr},
  publisher = {American Physical Society},
  doi = {10.1103/PhysRevLett.102.143601},
  url = {https://link.aps.org/doi/10.1103/PhysRevLett.102.143601}
}

@article{mss2014,
  title = {Cooperative Lamb Shift in a Mesoscopic Atomic Array},
  author = {Meir, Z. and Schwartz, O. and Shahmoon, E. and Oron, D. and Ozeri, R.},
  journal = {Phys. Rev. Lett.},
  volume = {113},
  issue = {19},
  pages = {193002},
  numpages = {5},
  year = {2014},
  month = {Nov},
  publisher = {American Physical Society},
  doi = {10.1103/PhysRevLett.113.193002},
  url = {https://link.aps.org/doi/10.1103/PhysRevLett.113.193002}
}

@article{pbj2014,
  title = {Observation of Suppression of Light Scattering Induced by Dipole-Dipole Interactions in a Cold-Atom Ensemble},
  author = {Pellegrino, J. and Bourgain, R. and Jennewein, S. and Sortais, Y. R. P. and Browaeys, A. and Jenkins, S. D. and Ruostekoski, J.},
  journal = {Phys. Rev. Lett.},
  volume = {113},
  issue = {13},
  pages = {133602},
  numpages = {5},
  year = {2014},
  month = {Sep},
  publisher = {American Physical Society},
  doi = {10.1103/PhysRevLett.113.133602},
  url = {https://link.aps.org/doi/10.1103/PhysRevLett.113.133602}
}

@article{bbl2016,
	doi = {10.1088/0953-4075/49/15/152001},
	url = {https://doi.org/10.1088/0953-4075/49/15/152001},
	year = 2016,
	month = {jun},
	publisher = {{IOP} Publishing},
	volume = {49},
	number = {15},
	pages = {152001},
	author = {Antoine Browaeys and Daniel Barredo and Thierry Lahaye},
	title = {Experimental investigations of dipole{\textendash}dipole interactions between a few {R}ydberg atoms},
	journal = {J. Phys. B: At. Mol. Opt. Phys.}
}

@Article{bzb2016,
author={Bromley, S. L.
and Zhu, B.
and Bishof, M.
and Zhang, X.
and Bothwell, T.
and Schachenmayer, J.
and Nicholson, T. L.
and Kaiser, R.
and Yelin, S. F.
and Lukin, M. D.
and Rey, A. M.
and Ye, J.},
title={Collective atomic scattering and motional effects in a dense coherent medium},
journal={Nat. Commun.},
year={2016},
month={Mar},
day={17},
volume={7},
number={1},
pages={11039},
issn={2041-1723},
doi={10.1038/ncomms11039},
url={https://doi.org/10.1038/ncomms11039}
}

@article{psr2017,
  title = {Cavity Antiresonance Spectroscopy of Dipole Coupled Subradiant Arrays},
  author = {Plankensteiner, David and Sommer, Christian and Ritsch, Helmut and Genes, Claudiu},
  journal = {Phys. Rev. Lett.},
  volume = {119},
  issue = {9},
  pages = {093601},
  numpages = {6},
  year = {2017},
  month = {Aug},
  publisher = {American Physical Society},
  doi = {10.1103/PhysRevLett.119.093601},
  url = {https://link.aps.org/doi/10.1103/PhysRevLett.119.093601}
}

@article{jbs2018,
  title = {Coherent scattering of near-resonant light by a dense, microscopic cloud of cold two-level atoms: Experiment versus theory},
  author = {Jennewein, Stephan and Brossard, Ludovic and Sortais, Yvan R. P. and Browaeys, Antoine and Cheinet, Patrick and Robert, Jacques and Pillet, Pierre},
  journal = {Phys. Rev. A},
  volume = {97},
  issue = {5},
  pages = {053816},
  numpages = {5},
  year = {2018},
  month = {May},
  publisher = {American Physical Society},
  doi = {10.1103/PhysRevA.97.053816},
  url = {https://link.aps.org/doi/10.1103/PhysRevA.97.053816}
}

@article{wzc2015,
  title = {Coherent Addressing of Individual Neutral Atoms in a 3D Optical Lattice},
  author = {Wang, Yang and Zhang, Xianli and Corcovilos, Theodore A. and Kumar, Aishwarya and Weiss, David S.},
  journal = {Phys. Rev. Lett.},
  volume = {115},
  issue = {4},
  pages = {043003},
  numpages = {5},
  year = {2015},
  month = {Jul},
  publisher = {American Physical Society},
  doi = {10.1103/PhysRevLett.115.043003},
  url = {https://link.aps.org/doi/10.1103/PhysRevLett.115.043003}
}

@article{chs2003,
  title = {Collective spontaneous emission from a line of atoms},
  author = {Clemens, J. P. and Horvath, L. and Sanders, B. C. and Carmichael, H. J.},
  journal = {Phys. Rev. A},
  volume = {68},
  issue = {2},
  pages = {023809},
  numpages = {19},
  year = {2003},
  month = {Aug},
  publisher = {American Physical Society},
  doi = {10.1103/PhysRevA.68.023809},
  url = {https://link.aps.org/doi/10.1103/PhysRevA.68.023809}
}

@Article{ymg2013,
author={Yan, Bo
and Moses, Steven A.
and Gadway, Bryce
and Covey, Jacob P.
and Hazzard, Kaden R. A.
and Rey, Ana Maria
and Jin, Deborah S.
and Ye, Jun},
title={Observation of dipolar spin-exchange interactions with lattice-confined polar molecules},
journal={Nature},
year={2013},
month={Sep},
day={01},
volume={501},
number={7468},
pages={521-525},
issn={1476-4687},
doi={10.1038/nature12483},
url={https://doi.org/10.1038/nature12483}
}

@article{mmm2018,
  title = {Lasing and Amplification from Two-Dimensional Atom Arrays},
  author = {Mkhitaryan, Vahagn and Meng, Lijun and Marini, Andrea and de Abajo, F. Javier Garc\'{\i}a},
  journal = {Phys. Rev. Lett.},
  volume = {121},
  issue = {16},
  pages = {163602},
  numpages = {5},
  year = {2018},
  month = {Oct},
  publisher = {American Physical Society},
  doi = {10.1103/PhysRevLett.121.163602},
  url = {https://link.aps.org/doi/10.1103/PhysRevLett.121.163602}
}

@article{jrp2017,
  title = {Many-Body Subradiant Excitations in Metamaterial Arrays: Experiment and Theory},
  author = {Jenkins, Stewart D. and Ruostekoski, Janne and Papasimakis, Nikitas and Savo, Salvatore and Zheludev, Nikolay I.},
  journal = {Phys. Rev. Lett.},
  volume = {119},
  issue = {5},
  pages = {053901},
  numpages = {6},
  year = {2017},
  month = {Aug},
  publisher = {American Physical Society},
  doi = {10.1103/PhysRevLett.119.053901},
  url = {https://link.aps.org/doi/10.1103/PhysRevLett.119.053901}
}

@Article{bpp2020,
author={Bekenstein, R.
and Pikovski, I.
and Pichler, H.
and Shahmoon, E.
and Yelin, S. F.
and Lukin, M. D.},
title={Quantum metasurfaces with atom arrays},
journal={Nature Physics},
year={2020},
month={Jun},
day={01},
volume={16},
number={6},
pages={676-681},
issn={1745-2481},
doi={10.1038/s41567-020-0845-5},
url={https://doi.org/10.1038/s41567-020-0845-5}
}

@article{poddubny1_2019,
  title = {Inelastic Scattering of Photon Pairs in Qubit Arrays with Subradiant States},
  author = {Ke, Yongguan and Poshakinskiy, Alexander V. and Lee, Chaohong and Kivshar, Yuri S. and Poddubny, Alexander N.},
  journal = {Phys. Rev. Lett.},
  volume = {123},
  issue = {25},
  pages = {253601},
  numpages = {6},
  year = {2019},
  month = {Dec},
  publisher = {American Physical Society},
  doi = {10.1103/PhysRevLett.123.253601},
  url = {https://link.aps.org/doi/10.1103/PhysRevLett.123.253601}
}

@Article{ma2022,
author={Masson, Stuart J.
and Asenjo-Garcia, Ana},
title={Universality of Dicke superradiance in arrays of quantum emitters},
journal={Nat. Commun.},
year={2022},
month={Apr},
day={27},
volume={13},
number={1},
pages={2285},
issn={2041-1723},
doi={10.1038/s41467-022-29805-4},
url={https://doi.org/10.1038/s41467-022-29805-4}
}

@article{FR2021,
  title = {Theoretical study of early-time superradiance for atom clouds and arrays},
  author = {Robicheaux, F.},
  journal = {Phys. Rev. A},
  volume = {104},
  issue = {6},
  pages = {063706},
  numpages = {10},
  year = {2021},
  month = {Dec},
  publisher = {American Physical Society},
  doi = {10.1103/PhysRevA.104.063706},
  url = {https://link.aps.org/doi/10.1103/PhysRevA.104.063706}
}

@Article{cv2014,
author={Chang, Darrick E.
and Vuleti{\'{c}}, Vladan
and Lukin, Mikhail D.},
title={Quantum nonlinear optics --- photon by photon},
journal={Nat. Photon.},
year={2014},
month={Sep},
day={01},
volume={8},
number={9},
pages={685-694},
issn={1749-4893},
doi={10.1038/nphoton.2014.192},
url={https://doi.org/10.1038/nphoton.2014.192}
}

@Article{ph2020,
author={Prasad, Adarsh S.
and Hinney, Jakob
and Mahmoodian, Sahand
and Hammerer, Klemens
and Rind, Samuel
and Schneeweiss, Philipp
and S{\o}rensen, Anders S.
and Volz, J{\"u}rgen
and Rauschenbeutel, Arno},
title={Correlating photons using the collective nonlinear response of atoms weakly coupled to an optical mode},
journal={Nat. Photon.},
year={2020},
month={Dec},
day={01},
volume={14},
number={12},
pages={719-722},
issn={1749-4893},
doi={10.1038/s41566-020-0692-z},
url={https://doi.org/10.1038/s41566-020-0692-z}
}

@article{rw2017,
  title = {Colloquium: Strongly interacting photons in one-dimensional continuum},
  author = {Roy, Dibyendu and Wilson, C. M. and Firstenberg, Ofer},
  journal = {Rev. Mod. Phys.},
  volume = {89},
  issue = {2},
  pages = {021001},
  numpages = {23},
  year = {2017},
  month = {May},
  publisher = {American Physical Society},
  doi = {10.1103/RevModPhys.89.021001},
  url = {https://link.aps.org/doi/10.1103/RevModPhys.89.021001}
}

@article{sp2023,
  title = {Waveguide quantum electrodynamics: Collective radiance and photon-photon correlations},
  author = {Sheremet, Alexandra S. and Petrov, Mihail I. and Iorsh, Ivan V. and Poshakinskiy, Alexander V. and Poddubny, Alexander N.},
  journal = {Rev. Mod. Phys.},
  volume = {95},
  issue = {1},
  pages = {015002},
  numpages = {59},
  year = {2023},
  month = {Mar},
  publisher = {American Physical Society},
  doi = {10.1103/RevModPhys.95.015002},
  url = {https://link.aps.org/doi/10.1103/RevModPhys.95.015002}
}

@article{lc2012,
author = {Liang, Yi and Czarnecki, Andrzej},
title = {Photon–photon scattering: a tutorial},
journal = {Canadian Journal of Physics},
volume = {90},
number = {1},
pages = {11-16},
year = {2012},
doi = {10.1139/p11-144},
URL = {https://doi.org/10.1139/p11-144},
}

@article{yr2008,
  title = {Multiphoton scattering in a one-dimensional waveguide with resonant atoms},
  author = {Yudson, V. I. and Reineker, P.},
  journal = {Phys. Rev. A},
  volume = {78},
  issue = {5},
  pages = {052713},
  numpages = {10},
  year = {2008},
  month = {Nov},
  publisher = {American Physical Society},
  doi = {10.1103/PhysRevA.78.052713},
  url = {https://link.aps.org/doi/10.1103/PhysRevA.78.052713}
}

@article{zb2013,
  title = {Persistent Quantum Beats and Long-Distance Entanglement from Waveguide-Mediated Interactions},
  author = {Zheng, Huaixiu and Baranger, Harold U.},
  journal = {Phys. Rev. Lett.},
  volume = {110},
  issue = {11},
  pages = {113601},
  numpages = {5},
  year = {2013},
  month = {Mar},
  publisher = {American Physical Society},
  doi = {10.1103/PhysRevLett.110.113601},
  url = {https://link.aps.org/doi/10.1103/PhysRevLett.110.113601}
}

@article{ah2019,
  title={Subradiant states of quantum bits coupled to a one-dimensional waveguide},
  author={Albrecht, Andreas and Henriet, Lo{\"\i}c and Asenjo-Garcia, Ana and Dieterle, Paul B and Painter, Oskar and Chang, Darrick E},
  journal={New J. Phys.},
  volume={21},
  number={2},
  pages={025003},
  year={2019},
  publisher={IOP Publishing}
}

@article{zm2019,
  title = {Theory of Subradiant States of a One-Dimensional Two-Level Atom Chain},
  author = {Zhang, Yu-Xiang and M\o{}lmer, Klaus},
  journal = {Phys. Rev. Lett.},
  volume = {122},
  issue = {20},
  pages = {203605},
  numpages = {5},
  year = {2019},
  month = {May},
  publisher = {American Physical Society},
  doi = {10.1103/PhysRevLett.122.203605},
  url = {https://link.aps.org/doi/10.1103/PhysRevLett.122.203605}
}

@article{zo2020,
  title = {Photon-Mediated Localization in Two-Level Qubit Arrays},
  author = {Zhong, Janet and Olekhno, Nikita A. and Ke, Yongguan and Poshakinskiy, Alexander V. and Lee, Chaohong and Kivshar, Yuri S. and Poddubny, Alexander N.},
  journal = {Phys. Rev. Lett.},
  volume = {124},
  issue = {9},
  pages = {093604},
  numpages = {6},
  year = {2020},
  month = {Mar},
  publisher = {American Physical Society},
  doi = {10.1103/PhysRevLett.124.093604},
  url = {https://link.aps.org/doi/10.1103/PhysRevLett.124.093604}
}

@article{zy2020,
  title = {Subradiant bound dimer excited states of emitter chains coupled to a one dimensional waveguide},
  author = {Zhang, Yu-Xiang and Yu, Chuan and M\o{}lmer, Klaus},
  journal = {Phys. Rev. Res.},
  volume = {2},
  issue = {1},
  pages = {013173},
  numpages = {9},
  year = {2020},
  month = {Feb},
  publisher = {American Physical Society},
  doi = {10.1103/PhysRevResearch.2.013173},
  url = {https://link.aps.org/doi/10.1103/PhysRevResearch.2.013173}
}

@article{p2020,
  title = {Quasiflat band enabling subradiant two-photon bound states},
  author = {Poddubny, Alexander N.},
  journal = {Phys. Rev. A},
  volume = {101},
  issue = {4},
  pages = {043845},
  numpages = {8},
  year = {2020},
  month = {Apr},
  publisher = {American Physical Society},
  doi = {10.1103/PhysRevA.101.043845},
  url = {https://link.aps.org/doi/10.1103/PhysRevA.101.043845}
}

@article{sf2007,
  title = {Strongly Correlated Two-Photon Transport in a One-Dimensional Waveguide Coupled to a Two-Level System},
  author = {Shen, Jung-Tsung and Fan, Shanhui},
  journal = {Phys. Rev. Lett.},
  volume = {98},
  issue = {15},
  pages = {153003},
  numpages = {4},
  year = {2007},
  month = {Apr},
  publisher = {American Physical Society},
  doi = {10.1103/PhysRevLett.98.153003},
  url = {https://link.aps.org/doi/10.1103/PhysRevLett.98.153003}
}

@Article{m2010,
author={Miller, David A. B.},
title={Are optical transistors the logical next step?},
journal={Nat. Photon.},
year={2010},
month={Jan},
day={01},
volume={4},
number={1},
pages={3-5},
issn={1749-4893},
doi={10.1038/nphoton.2009.240},
url={https://doi.org/10.1038/nphoton.2009.240}
}

@Article{k2008,
author={Kimble, H. J.},
title={The quantum internet},
journal={Nature},
year={2008},
month={Jun},
day={01},
volume={453},
number={7198},
pages={1023-1030},
issn={1476-4687},
doi={10.1038/nature07127},
url={https://doi.org/10.1038/nature07127}
}

@article{ms2004,
doi = {10.1088/1464-4266/6/6/017},
url = {https://dx.doi.org/10.1088/1464-4266/6/6/017},
year = {2004},
month = {may},
publisher = {},
volume = {6},
number = {6},
pages = {S575},
author = {Ashok Muthukrishnan and  Marlan O Scully and  M Suhail Zubairy},
title = {Quantum microscopy using photon correlations},
journal = {J. Opt. B: Quantum Semiclass. Opt.}
}

@Article{gl2011,
author={Giovannetti, Vittorio
and Lloyd, Seth
and Maccone, Lorenzo},
title={Advances in quantum metrology},
journal={Nat. Photon.},
year={2011},
month={Apr},
day={01},
volume={5},
number={4},
pages={222-229},
issn={1749-4893},
doi={10.1038/nphoton.2011.35},
url={https://doi.org/10.1038/nphoton.2011.35}
}

@article{ms2021,
  title = {Bound Photonic Pairs in 2D Waveguide Quantum Electrodynamics},
  author = {Marques, Y. and Shelykh, I. A. and Iorsh, I. V.},
  journal = {Phys. Rev. Lett.},
  volume = {127},
  issue = {27},
  pages = {273602},
  numpages = {6},
  year = {2021},
  month = {Dec},
  publisher = {American Physical Society},
  doi = {10.1103/PhysRevLett.127.273602},
  url = {https://link.aps.org/doi/10.1103/PhysRevLett.127.273602}
}

@article{pp2023,
  title = {Bound state of distant photons in waveguide quantum electrodynamics},
  author = {Poshakinskiy, Alexander V. and Poddubny, Alexander N.},
  journal = {Phys. Rev. A},
  volume = {108},
  issue = {2},
  pages = {023707},
  numpages = {7},
  year = {2023},
  month = {Aug},
  publisher = {American Physical Society},
  doi = {10.1103/PhysRevA.108.023707},
  url = {https://link.aps.org/doi/10.1103/PhysRevA.108.023707}
}

@article{uk2024,
  title = {Nonradiant multiphoton states in quantum ring oligomers},
  author = {Ustimenko, N. and Kornovan, D. and Volkov, I. and Sheremet, A. and Savelev, R. and Petrov, M.},
  journal = {Phys. Rev. A},
  volume = {110},
  issue = {1},
  pages = {L011501},
  numpages = {7},
  year = {2024},
  month = {Jul},
  publisher = {American Physical Society},
  doi = {10.1103/PhysRevA.110.L011501},
  url = {https://link.aps.org/doi/10.1103/PhysRevA.110.L011501}
}

@article{ff2023,
  title = {Optical control of collective states in one-dimensional ordered atomic chains beyond the linear regime},
  author = {Fayard, N. and Ferrier-Barbut, I. and Browaeys, A. and Greffet, J.-J.},
  journal = {Phys. Rev. A},
  volume = {108},
  issue = {2},
  pages = {023116},
  numpages = {12},
  year = {2023},
  month = {Aug},
  publisher = {American Physical Society},
  doi = {10.1103/PhysRevA.108.023116},
  url = {https://link.aps.org/doi/10.1103/PhysRevA.108.023116}
}

@article{je2024,
  title = {Tailoring the statistics of light emitted from two interacting quantum emitters},
  author = {Juan-Delgado, Adri\'an and Esteban, Ruben and Nodar, \'Alvaro and Trebbia, Jean-Baptiste and Lounis, Brahim and Aizpurua, Javier},
  journal = {Phys. Rev. Res.},
  volume = {6},
  issue = {2},
  pages = {023207},
  numpages = {26},
  year = {2024},
  month = {May},
  publisher = {American Physical Society},
  doi = {10.1103/PhysRevResearch.6.023207},
  url = {https://link.aps.org/doi/10.1103/PhysRevResearch.6.023207}
}

@article{wb2020,
  title = {Superatom Picture of Collective Nonclassical Light Emission and Dipole Blockade in Atom Arrays},
  author = {Williamson, L. A. and Borgh, M. O. and Ruostekoski, J.},
  journal = {Phys. Rev. Lett.},
  volume = {125},
  issue = {7},
  pages = {073602},
  numpages = {7},
  year = {2020},
  month = {Aug},
  publisher = {American Physical Society},
  doi = {10.1103/PhysRevLett.125.073602},
  url = {https://link.aps.org/doi/10.1103/PhysRevLett.125.073602}
}

@article{sc2024,
  title = {Coherent Control of Photon Correlations in Trapped Ion Crystals},
  author = {Singh, K. and Cidrim, A. and Kovalenko, A. and Pham, T. M. and \ifmmode \check{C}\else \v{C}\fi{}\'{\i}p, O. and Slodi\ifmmode \check{c}\else \v{c}\fi{}ka, L. and Bachelard, R.},
  journal = {Phys. Rev. Lett.},
  volume = {134},
  issue = {20},
  pages = {203602},
  numpages = {7},
  year = {2025},
  month = {May},
  publisher = {American Physical Society},
  doi = {10.1103/PhysRevLett.134.203602},
  url = {https://link.aps.org/doi/10.1103/PhysRevLett.134.203602}
}

@article{cs2023,
  title = {Tailoring Photon Statistics with an Atom-Based Two-Photon Interferometer},
  author = {Cordier, Martin and Schemmer, Max and Schneeweiss, Philipp and Volz, J\"urgen and Rauschenbeutel, Arno},
  journal = {Phys. Rev. Lett.},
  volume = {131},
  issue = {18},
  pages = {183601},
  numpages = {5},
  year = {2023},
  month = {Oct},
  publisher = {American Physical Society},
  doi = {10.1103/PhysRevLett.131.183601},
  url = {https://link.aps.org/doi/10.1103/PhysRevLett.131.183601}
}

@book{jackson,
    author = {Jackson, J. D.},
    title = {Classical Electrodynamics},
    publisher = {Wiley},
    edition = {3rd ed.},
    year = {1999}
}

@article{wrv2020,
  title = {Light of Two Atoms in Free Space: Bunching or Antibunching?},
  author = {Wolf, Sebastian and Richter, Stefan and von Zanthier, Joachim and Schmidt-Kaler, Ferdinand},
  journal = {Phys. Rev. Lett.},
  volume = {124},
  issue = {6},
  pages = {063603},
  numpages = {6},
  year = {2020},
  month = {Feb},
  publisher = {American Physical Society},
  doi = {10.1103/PhysRevLett.124.063603},
  url = {https://link.aps.org/doi/10.1103/PhysRevLett.124.063603}
}

@article{reset-matrix,
  title = {How to reset an atom after a photon detection: Applications to photon-counting processes},
  author = {Hegerfeldt, Gerhard C.},
  journal = {Phys. Rev. A},
  volume = {47},
  issue = {1},
  pages = {449--455},
  numpages = {0},
  year = {1993},
  month = {Jan},
  publisher = {American Physical Society},
  doi = {10.1103/PhysRevA.47.449},
  url = {https://link.aps.org/doi/10.1103/PhysRevA.47.449}
}

@article{agp2011,
  title={Few emitters in a cavity: from cooperative emission to individualization},
  author={Auff{\`e}ves, Alexia and Gerace, Dario and Portolan, Stefano and Drezet, Aur{\'e}lien and Santos, M Fran{\c{c}}a},
  journal={New J. Phys.},
  doi={10.1088/1367-2630/13/9/093020},
  volume={13},
  number={9},
  pages={093020},
  year={2011},
  publisher={IOP Publishing}
}

@article{tw2009,
author = {Vasily V. Temnov and Ulrike Woggon},
journal = {Opt. Express},
number = {7},
pages = {5774--5782},
publisher = {Optica Publishing Group},
title = {Photon statistics in the cooperative spontaneous emission},
volume = {17},
month = {Mar},
year = {2009},
url = {https://opg.optica.org/oe/abstract.cfm?URI=oe-17-7-5774},
doi = {10.1364/OE.17.005774},
}

@article{mh2010,
  title = {Intensity fluctuations in steady-state superradiance},
  author = {Meiser, D. and Holland, M. J.},
  journal = {Phys. Rev. A},
  volume = {81},
  issue = {6},
  pages = {063827},
  numpages = {7},
  year = {2010},
  month = {Jun},
  publisher = {American Physical Society},
  doi = {10.1103/PhysRevA.81.063827},
  url = {https://link.aps.org/doi/10.1103/PhysRevA.81.063827}
}

@article{gsd2005,
  title = {Photon Statistics from Coupled Quantum Dots},
  author = {Gerardot, Brian D. and Strauf, Stefan and de Dood, Michiel J. A. and Bychkov, Andrey M. and Badolato, Antonio and Hennessy, Kevin and Hu, Evelyn L. and Bouwmeester, Dirk and Petroff, Pierre M.},
  journal = {Phys. Rev. Lett.},
  volume = {95},
  issue = {13},
  pages = {137403},
  numpages = {4},
  year = {2005},
  month = {Sep},
  publisher = {American Physical Society},
  doi = {10.1103/PhysRevLett.95.137403},
  url = {https://link.aps.org/doi/10.1103/PhysRevLett.95.137403}
}

@article{r2023,
  title = {Cooperative quantum-optical planar arrays of atoms},
  author = {Ruostekoski, Janne},
  journal = {Phys. Rev. A},
  volume = {108},
  issue = {3},
  pages = {030101},
  numpages = {33},
  year = {2023},
  month = {Sep},
  publisher = {American Physical Society},
  doi = {10.1103/PhysRevA.108.030101},
  url = {https://link.aps.org/doi/10.1103/PhysRevA.108.030101}
}

@article{l1970,
  title = {Radiation from an $N$-Atom System. I. General Formalism},
  author = {Lehmberg, R. H.},
  journal = {Phys. Rev. A},
  volume = {2},
  issue = {3},
  pages = {883--888},
  numpages = {0},
  year = {1970},
  month = {Sep},
  publisher = {American Physical Society},
  doi = {10.1103/PhysRevA.2.883},
  url = {https://link.aps.org/doi/10.1103/PhysRevA.2.883}
}

@book{loudon,
    author = {Loudon, Rodney},
    title = {The Quantum Theory of Light},
    publisher = {Oxford University Press},
    year = {2000},
    month = {09},
    isbn = {9780198501770},
    doi = {10.1093/oso/9780198501770.001.0001},
    url = {https://doi.org/10.1093/oso/9780198501770.001.0001},
}

@article{zmw1987,
  title = {Quantum jumps in atomic systems},
  author = {Zoller, P. and Marte, M. and Walls, D. F.},
  journal = {Phys. Rev. A},
  volume = {35},
  issue = {1},
  pages = {198--207},
  numpages = {0},
  year = {1987},
  month = {Jan},
  publisher = {American Physical Society},
  doi = {10.1103/PhysRevA.35.198},
  url = {https://link.aps.org/doi/10.1103/PhysRevA.35.198}
}

@article{csv1989,
  title = {Photoelectron waiting times and atomic state reduction in resonance fluorescence},
  author = {Carmichael, H. J. and Singh, Surendra and Vyas, Reeta and Rice, P. R.},
  journal = {Phys. Rev. A},
  volume = {39},
  issue = {3},
  pages = {1200--1218},
  numpages = {0},
  year = {1989},
  month = {Feb},
  publisher = {American Physical Society},
  doi = {10.1103/PhysRevA.39.1200},
  url = {https://link.aps.org/doi/10.1103/PhysRevA.39.1200}
}
 
\end{document}